\pdfoutput=1
\documentclass[epj,final]{svjour}

\usepackage{latexsym}
\usepackage{url}
\usepackage{amsfonts}
\usepackage{longtable}

\RequirePackage{graphicx}

\begin{document}
\title{The light speed vs the observer: the Kennedy-Thorndike test from GRAAL-ESRF}
\author{V.G. Gurzadyan\inst{1,2} \and A.T. Margaryan\inst{1}
}                     
%
%
\institute{Center for Cosmology and Astrophysics, Alikhanian National Laboratory and Yerevan State University, Yerevan, Armenia  \and SIA, Sapienza Universita di Roma, Rome, Italy 
}

\date{Received: date / Revised version: date}
%

\abstract{High precision tests of the light speed constancy for all observers as of empirical basis of the Special Relativity have continuously been among the goals of advanced experimental studies. Based on the Compton Edge method proposed by us \cite{GM}, a constraint on the one-way light speed isotropy and the Lorentz invariance violation has been obtained at the dedicated GRAAL experiment at European Synchrotron Radiation Facility (ESRF, Grenoble) \cite{GR2005,GR2007,GR2010,GR2012}. Using the GRAAL's data we now get a new constraint on one of key tests of Special Relativity - the Kennedy-Thorndike experiment  \cite{KT}- in probing the light speed invariance with respect to the velocity of the observer (apparatus). Our analysis takes advantage of GRAAL's setup where two separate energy scales are involved: first, via the position of the Compton Edge determining the light speed in the reference frame of incident 6 GeV electrons within the tagging system, second, in the calorimeter via the 1.27 MeV photons of the $^{22}$Na source. The two energy scales are engaged to each other through production of $\eta$ mesons by tagged laser Compton backscattered $\gamma$-rays. Accuracy of the calibration and stability of energies reached in each section enable us to obtain the limit $7\, 10^{-12}$ for the Kennedy-Thorndike test, which improves the currently existing limits by three orders of magnitude.   
} 

\maketitle
\section*{Introduction}

The light speed constancy for all observers and the equivalence principle are the key empirical bases of Einstein's Special and General Relativities. Long after the creation of Relativity theories Einstein remained attentive to the ongoing experiments on the empirical bases \cite{E}. The observational evidence for the existence of dark energy and dark matter have increased the interest to models with varying light speed and Lorentz invariance violation (LIV) and hence to relevant experimental activity. Particularly, the tests of the light speed invariance with respect to the direction (isotropy) and the LIV models have always been among the goals of high precision experiments \cite{Mat,Kos,Col}. Their link to the General Relativity and gravitational waves gained further interest upon the LIGO-Virgo's detection of gravitation waves \cite{Shoe}, while recent satellite studies enable one to improve the equivalence principle precision limits.  
 
Three types of light speed involving experiments have been considered of particular interest (see \cite{Lam} and references therein), although each measuring a particular effect, they are mutually complementary in probing the Special Relativity: 

(a) Light speed isotropy, i.e. invariance to the direction (Michelson-Morley (MM) test);

(b) Light speed independence on the velocity of the observer (Kennedy-Thorndike (KT) test) \cite{KT};

(c) Time delation (Ives and Stilwell (IS) test). 
 
Among the experiments on the MM-test there were the measurements based on the idea of the daily monitoring of the Compton Edge (CE) which corresponds to the maximal energy of the scattered photons (see below), with respect to the frame of cosmic microwave background (CMB) as suggested in \cite{GM}. That idea has been realized at GRAAL experiment of the European Synchrotron Radiation Facility (ESRF, Grenoble) at the scattering of accelerated electrons of 6.03 GeV energy and laser monochromatic photons. The results of those studies enabled to constrain the one-way light speed isotropy to the precision of 10$^{-14}$ (MM-test) \cite{GR2005,GR2007,GR2010,GR2012}.

Here we use the data of the GRAAL experiment to obtain a constraint also to the other key test, namely, the Kennedy-Thorndike one \cite{KT}. 
The current KT-test precision is $10^{-7}-10^{-8}$ \cite{Wo,To}, for details and references we refer to reviews \cite{Lam,Mat,Kos}; among the proposed KT-tests is a dedicated satellite experiment \cite{Li}. 

More specifically, to probe the light speed invariance with respect to the velocity of the apparatus, namely, the with respect to the beam electron reference system and the reference system of the calorimeter, we use the data of GRAAL's Laser Compton Backscattered (LCB) experiment, the energy calibration accuracy and stability of the Compton Edge obtained in the tagging system and of the $\eta$-meson production detected in the BGO electromagnetic calorimeter  \cite{graal98,graal07}.
   
Three issues make efficient the use of the GRAAL data for the KT-test:
 
 (a) the $\gamma^2$-dependence ($\gamma$ is the Lorentz factor) of energy of the Compton Edge;
         
 (b) the calibration of the absolute energy scale of the internal tagging system by CE;
 
 (c) the calibration of the absolute energy scale of the electromagnetic calorimeter using the 1.27 MeV photons from $^{22}$Na source.
 
In the analysis below we obtain the accuracy of the invariance of the speed of light via its evaluation, first, with respect to the frame of the incident electron undergoing the Compton scattering,  second, to the laboratory frame in the calorimeter. 
We obtain the limit $7.1 \, 10^{-12}$ for the KT-test - on the light speed invariance with respect to the velocity of the apparatus - which is better than the existing limits by 3 orders of magnitude.    

\section{Compton Edge}

The kinematics of the Compton scattering defines the energy $\omega_s$ of photons having scattered on the electrons of energy $E_e$  in dependence on the energy of the primary photon $\omega_0$ as follows (e.g. \cite{GM})
\begin{equation}
\omega_s =\frac{(1- \beta cos\theta) \omega_{0}}{1 - \beta cos \theta_{\gamma} + (1- cos \theta_0)(\omega_{0}/E_e)},
\end{equation}
where $\theta_0$ is the angle between incident and scattered photons' momenta, and $\theta$, $\theta_{\gamma}$ are the angles between the momentum of electron and the incident and the scattered photons, respectively, and the Lorentz factor of electron in the laboratory frame is
\begin{equation}
\gamma=(1-\beta^2)^{-1/2},
\end{equation}
and  $\beta= v/c$, $v$ is electron velocity.

From Eq.(1) the maximum energy of the scattered photons reached at small angle limit, $\theta_{\gamma}\to 0$,  called Compton Edge, and at head-on collision $\theta_0,\theta = \pi$, yields
\begin{equation}
\omega_{CE}= \frac{(1+\beta)^2\gamma^2 \omega_0}{1+\frac{2(1+\beta)\gamma \omega_0}{mc^2}},
\end{equation}
where $m$ is electron mass. 

When the electron beam energy is kept stable to high accuracy, then from Eq.(2) 
\begin{equation}
\beta d\beta = \frac{1}{\gamma^2} \frac{d\gamma}{\gamma}.
\end{equation}  

The Compton scattering of laser photons of energy $\omega_0$ and high energy electrons of Lorentz factor $\gamma$ can be represented as a four step process, if one accepts the potential possibility of non-constancy of the light speed:

1. In the laboratory frame the photons of energy $\omega_0$ travel at speed $c_1$ (allowing speed anisotropy), the electrons possess energies $E_1$ and velocity $v$;

2. In the rest frame of initial electrons the Doppler shifted photons (due to relativistic time dilation) of energy 
\begin{equation}
\omega_{01}=\gamma_2 (1 - \beta_2 cos\theta)\omega_0
\end{equation} 
scatter over electrons at velocity $c_2$ (allowing possible velocity dependence in the frame of the electron), were $\beta_2=v/c_2$. In the case of head-on collisions $\theta=\pi$ and $\omega_{e1}=\gamma_2 (1+\beta_2)\omega_0$. 

3. In the rest frame of an electron, 180$^\circ$ Compton scattered photons of energies 
\begin{equation}
\omega_{02}=\omega_{e1}/(1+2\omega_{e1}/mc^2)
\end{equation}
move away at velocity $c_3$ (due to photon direction change by 180$^\circ$). The recoil electrons have equal and opposite momenta.

4. Finally, in the laboratory frame we have the Compton scattered and Doppler shifted laser photons of energies 
\begin{equation}
\omega_{21}=\gamma_4 (1+\beta_4 cos \theta_{\gamma})\omega_{02}
\end{equation}
and velocity $c_4$ (allowing direction dependence), where 
$\beta_4=v/c_4$, $\gamma_4 =(1-\beta^2_4)^{-1/2}$  and recoil electrons of energies $E_2=E_1 -\omega_{21}$. In our case $\theta_{\gamma}=0$,  $\theta_0,\theta = \pi$ and for CE we have 
\begin{equation}
\omega^{max}_{21}=\gamma_4(1+\beta_4)\omega_{02}=\gamma_2 \gamma_4(1+\beta_4)^2 \omega_{01}/(1+2\gamma_4(1+\beta_4)\omega_{01}/mc^2)
\end{equation}
and 
\begin{equation}
E^{min}_2=E_1-\omega^{max}_{21}.
\end{equation}

In Special Relativity, obviously, $c_i=c, \beta_i=\beta,\gamma_i=\gamma$, whence the $\omega^{max}_{21}$ in Eq.(9) coincides with $\omega_{CE}$ in Eq.(3), thus confirming that we deal with the light speed in the initial electron's frame.

\section{The Compton Edge at GRAAL: the light speed in the moving electron's frame}

The GRAAL facility installed in ESRF involved a $\gamma$-ray beam originated at Compton scattering of 514 nm and 351 nm laser photons and 6.03 GeV electrons accelerated in the storage ring \cite{Bo,graal98,graal07}. The incident photons were produced by a high-power Ar laser located in about 40 m from the intersection location in the tagging box. The laser beam entered the vacuum region through a MgF window and then by Al-coated Be mirror was directed towards the electron beam. The laser photon and electron beams overlapped within 6.5 m long section. Then the scattered photons were absorbed within a four-quadrant calorimeter, which enabled one to stabilize the center of the laser beam within 0.1 mm. The scattered electrons were extracted from the main beam by means of a magnetic dipole located after the straight section. 

\begin{figure}[!htbp]
  \centering
  \includegraphics[width=100mm]{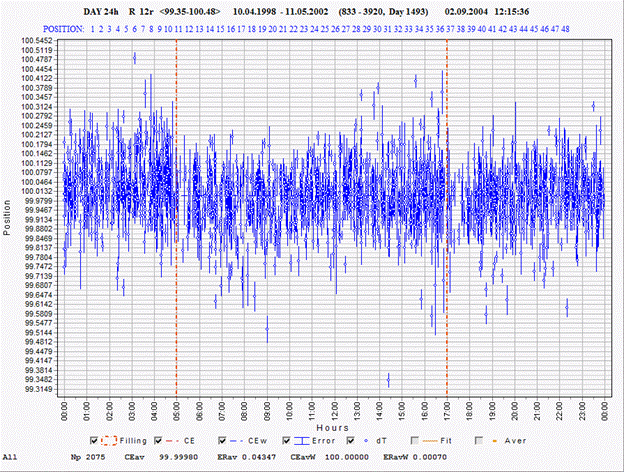}
  \caption{GRAAL data: CE vs 24-hour period; red vertical lines denote the beam refill instances \cite{GR2005,GR2010}.}
\end{figure}

\begin{figure}[!htbp]
  \centering
  \includegraphics[width=100mm]{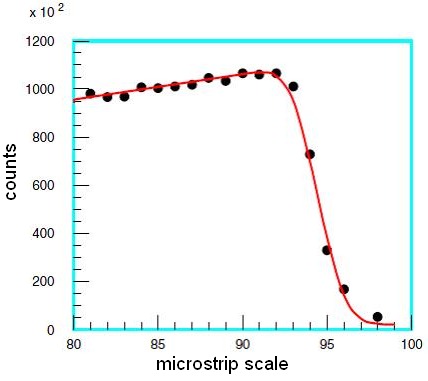}
  \caption{The Compton Edge for the green laser line (2.41 eV) observed at the GRAAL facility; the counts vs the microstrip detector scale, details in \cite{GR2005}.}
\end{figure}

The position of the electron beam could then be accurately measured within the tagging system located in 50 cm from the exit of the dipole. 
The tagging system played the role of a magnetic spectrometer which enabled one to extract the information on the scattered electron (photon) momenta. The tagging system included a position-sensitive Si $\mu$-strip detector of 128 strips of 300 $\mu m$ pitch and 500 $\mu m$ thickness each, linked to a set of fast plastic scintillators for the timing information and triggering the data acquisition. The detectors were located in a movable box shielded for a strong X-ray background generated in the dipole. The X-ray background led to a remarkable heat release, which led to temperature variations inside the tagging box correlated with the ESRF beam intensity. This effect was controlled and corrected.

Eq.(3) for the CE in the case of the GRAAL setup has the form (see \cite{GR2005}) 
\begin{equation}
\omega_{CE}= \frac{\gamma m_e X_{CE}}{A+X_{CE}},
\end{equation}
where $A$ is a constant, $X_{CE}$ is the distance of the scattered electron's position in the tagging detector from the initial beam.

Then, from Eqs.(3),(10) one has
\begin{equation}
X_{CE}= \frac{4\gamma A \omega_0}{m}
\end{equation}
and hence
\begin{equation}
\frac{\delta X_{CE}}{X_{CE}}=\frac{\delta \gamma}{\gamma} + \frac{\delta A}{A} + \frac{\delta \omega_0}{\omega_0}.
\end{equation}
Then, as follows from Eqs.(4),(10-12), the position of the CE defines the light speed variation \cite{GM,GR2005}
\begin{equation}
\frac{\delta c}{c}= \gamma^{-2} \frac{\delta X_{CE}}{X_{CE}}.
\end{equation}
Fig. 1 exhibits the GRAAL's CE data (2075 points) vs the 24 hour period; the vertical axis is $X_{CE}$ in GRAAL's microstrip scale units. The MM-test was obtained via such monitoring for daily variations of CE, for details see \cite{GR2005,GR2010}. For MM-test the crucial was the stability of the beam energy and tagging system, while the stability of the thermal expansion of the tagging system was a subject of a separate study (for details see \cite{GR2010} and also the discussion in \cite{jlab}). That resulted in relative CE accuracy and enabled one to constrain the light speed direction-dependence precision for the Lorentz factor of the ESRF electron beam $\gamma = 11820$ to an accuracy \cite{GR2010,GR2012} 
\begin{equation}
\delta c(\theta)/c \simeq 10^{-14}. 
\end{equation}
For KT-test, however, the precision not of relative but of an absolute value of the energy scales are needed, which as we show below, is also possible to involve both due to the experimental setup of the GRAAL facility and the high stability of the relevant parameters. 

The CE method, i.e. the precise evaluation of CE position (Fig. 2) has been efficiently used at GRAAL experiment in order to calibrate the tagging system. Namely, the CE has been localized at precision
about 10 $\mu m$ of the 128-strip tagging box which corresponded to  \cite{graal07}
\begin{equation}
\frac{\Delta E_{\gamma}}{E_{\gamma}} \simeq 2\, 10^{-4}.
\end{equation}  
The stability and high resolution of the accelerator electron beam and GRAAL's tagging system allow to achieve the accuracy for the light speed isotropy limit $10^{-14}$. In this way the energy scale of the tagging system is determined by the light speed in the incident electron's frame.  

\section{The BGO calorimeter: the light speed in the laboratory frame}

A high resolution and large solid angle BGO electromagnetic calorimeter combined with multiwire proportional chambers and scintillator counters was another essential section of the GRAAL facility \cite{graal98}. The calorimeter made of 480 crystals each of 21 radiation
lengths was used for detection of $\gamma$-photons coming e.g. from $\eta \to 2\gamma$ decay.
The crystals were distributed within clusters and the photon energy resolution was about 3\%.
For 3 cm target the angular resolution yielded 6$^\circ$ and 7$^{\circ}$ for polar and azimuthal angles, respectively.

The absolute calibration of crystals and its monitoring has been performed using the 1.27 MeV photons from a $^{22}$Na source
and via creation of a special monitoring system \cite{graal98,graal07}. The nonlinearity of this setup which included the electronics
and BGO detectors is on the level of $10^{-3}$.  Thus, in this approach involving $^{22}$Na the energy scale of the BGO calorimeter is determined by the light speed in the laboratory frame.

The recoil proton track is measured by a set of Multi-Wire Proportional Chambers (MWPC) with the average polar and azimuthal resolutions of 1.5 and 2 degrees, respectively, for the forward angle tracks. For the charged particles emitted in the forward direction, a Time-of-Flight (ToF) measurement is provided by a double scintillator telescope placed at a distance of 3 m from the target and having a resolution of ~600 ps, the calibration of which was obtained from fast electrons produced in the target.

The GRAAL experiment has traced the reaction  $\gamma + p \to \eta + p$. Only events with two neutral clusters in BGO calorimeter and a single charged particle track were selected. Channel selection was achieved by applying the following cuts on the invariant mass of detected photons ($M_{2g}$); the energy of $\eta$ mesons ($R_\eta = E_\eta/E^*_\eta$); the direction of protons ($d\theta_p=\theta^*_p - \theta_p, \, d\phi_p=\phi^*_p - \phi_p$); the energy of protons ($dt_p = ToF^*_p - ToF_p$), where "*" denotes the variables calculated from the two-body kinematics as opposed to the measured ones, for details see \cite{graal07}. The applied cuts and two-body kinematics directly relate the invariant and missing masses of $\eta$-mesons to the energies of incoming photons.   

In Fig.3ab two examples of experimental distributions \cite{graal07} with the invariant mass of the $\eta$ and the missing mass calculated from the recoil proton momentum along with their fits are given. We have computed the 9th order polynomial fit for both data in Fig.3ab. The results in Fig.3a determine the mean value of the invariant mass of the $\eta$ meson as $m_{\eta}= 548.46$ MeV with standard error $0.05$ MeV for about $10^6$ events; the computed mean value differs from its PDG value $m^a_{\eta}= 547.862 \pm 0.17$ MeV \cite{Am} as $548.46 - 547.86=0.6$ i.e. to an accuracy $1.1\, 10^{-3}$. For Fig.3b we similarly obtain the $\eta$ missing mass $m_{\eta}= 552.16$  with standard error $0.058$ which differs from PDG value as $552.16 - 547.86=4.3$ i.e. to an accuracy $7.8\, 10^{-3}$. This less precision is obviously due to ToF system calibration, which is used for proton momentum determination.  It is worthy to mention that, the data obtained without CE calibration of the internal tagging system define the $\eta$ meson invariant mass as 542.8 MeV, i.e. to a lower accuracy $10^{-2}$ \cite{graal98}.

\begin{figure}[!htbp]
  \centering
  \includegraphics[width=100mm]{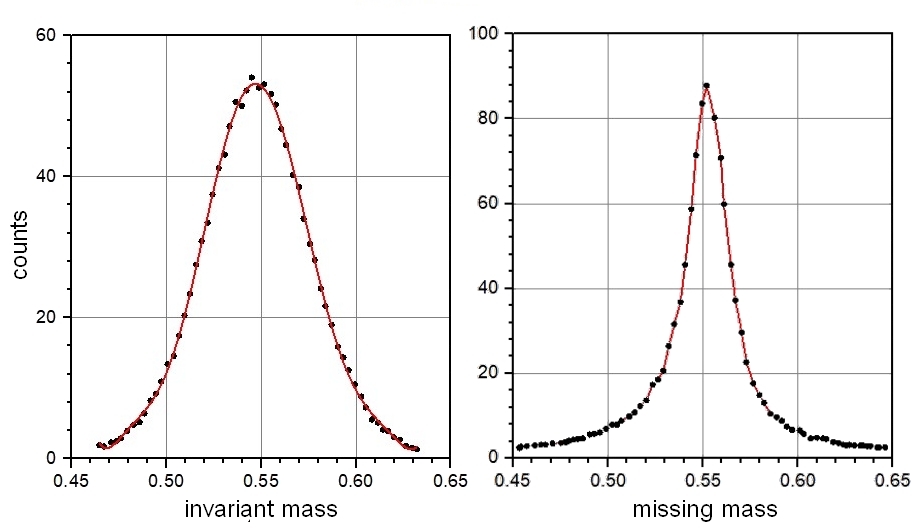}
  \caption{(a) Invariant mass spectrum for $\eta \to  2\gamma$ and (b) missing
mass spectrum calculated from the proton momentum, the data (dotes) from \cite{graal07}.  The best fit curves for both data
  are shown; the axes denote the counts (in arbitrary units) vs the mass scale (in $GeV/c^2$).}
\end{figure}

\section{The KT-test at GRAAL}

We are interested in the use of the above mentioned results of the GRAAL experiment for the K-T test (see also \cite{jlab}). The idea is the following. 
Two processes can be distinguished at the GRAAL's experiment which respectively define two energy scales for photons i.e. of the LCB $\gamma$-rays and of the BGO calorimeter. The first one is the CE position which is determined by the light speed in the moving electron's frame, the second one is the BGO calorimeter energy scale determined by the 1.27 MeV photons of $^{22}$Na. 

The two separate energy scales of the GRAAL experiment are linked to each other through the $\eta$-meson production process by Compton backscattered photons.  Then, from the accuracy of the measured $\eta$-meson invariant mass, i.e. from the fit of the data in Fig.3a we have  
\begin{equation}
\frac{d \gamma}{\gamma} = \frac{m_{\eta} - m^a_{\eta}}{m_{\eta}} \leq 1.1 \, 10^{-3}. 
\end{equation}  

It is remarkable, that this accuracy is reached from the energy calibration using the CE position with a precision about $10 \mu m$ or $\Delta E_{\gamma}/E_{\gamma} \simeq 2\, 10^{-4}$ \cite{graal07} and taking into account nonlinearities of the BGO calorimeter on the level of $10^{-3}$ \cite{graal98}:  the energy calibration of the tagged system is extracted run by run from the fit of the CE position with that precision. 
From here, in view of Eqs.(13), (16), we arrive at the light speed constancy in the incident accelerator electron's frame to accuracy   
\begin{equation}
\delta c(v)/ c  = 7.1 \, 10^{-12}.
\end{equation}  

This defines the precision of the KT-test as of light speed's invariance with respect to the velocity $v$ of the observer's frame, i.e. of the electron of that velocity vs the laboratory one.

\section{Conclusions}

Kennedy-Thorndike experiment is attributed to one of main tests for the Special Relativity and the Lorentz invariance. That test concerns the light speed invariance with respect to the velocity of the inertial frame of the observer. The CE data of GRAAL-ESRF experiment which previously have been efficiently used for Michelson-Morley-like test, i.e. for probing the direction-dependence of the light speed via  monitoring of the stability of CE, are now shown to be informative also for KT-test. In the performed analysis the invariance of the light speed is tested with respect to the accelerated 6 GeV electron's frame and to the laboratory's one.  The obtained accuracy is better than existing KT-limits by three orders of magnitude and yields  $7.1 \, 10^{-12}$. 

This confirms the power of the Compton Edge method for revealing information on the basic physics in accelerator experiments, as well as the advanced parameters reached at ESRF and GRAAL's experimental setup. 

We are thankful to J.-P. Bocquet, C. Schaerf and D. Rebreyend of the GRAAL collaboration and A. Kashin for numerous discussions.

\end{document}